\documentclass[preprint,12pt,authoryear]{elsarticle}

\usepackage{amssymb}
\usepackage[labelfont=bf]{caption}
\usepackage{amsmath}
\usepackage{float}
\usepackage{url}
\usepackage{epsfig}

\journal{Journal of Theoretical Biology}

\begin{document}
\newpageafter{author}
\begin{frontmatter}



\title{Coupled opinion-environmental dynamics in polarized and prejudiced populations} 

\author{Cameron Kerr\corref{cor1}\fnref{label1}} 
\author{Madhur Anand\fnref{label2}}
\author{Chris T. Bauch\fnref{label1}}
\cortext[cor1]{Corresponding author, \textit{E-mail address}: c5kerr@uwaterloo.ca}

\affiliation[label1]{organization={Department of Applied Mathematics, University of Waterloo},
            addressline={200 University Ave W},
            city={Waterloo},
            postcode={N2L 3G1},
            state={Ontario},
            country={Canada}}
\affiliation[label2]{organization={School of Environmental Sciences, University of Guelph},
            addressline={50 Stone Rd E},
            city={Guelph},
            postcode={N1G 2W1},
            state={Ontario},
            country={Canada}}
\begin{abstract}
    Public opinion on environmental issues remains polarized in many countries, posing a significant barrier to the implementation of effective policies. Behind this polarization, empirical studies have identified social susceptibility, personal prejudice, and personal experience as dominant factors in opinion formation on environmental issues. However, current coupled human-environment models have not yet incorporated all three factors in polarized populations. We developed a stylized coupled human-environment model to investigate how social susceptibility, personal prejudice, and personal experience shape opinion formation and the environment in polarized populations. Using analytical and numerical methods, we characterized the conditions under which polarization, consensus, opinion changes, and cyclic dynamics emerge depending on the costs of mitigation, environmental damage, and the factors influencing opinion formation. Our model shows that prejudice is the key driver of persistent polarization, with even slightly prejudiced populations maintaining indefinite polarization independent of their level of objectivity. We predict that polarization can be reduced by decreasing the role of prejudice or increasing the willingness to consider opposing opinions. Finally, our model shows that cost reduction methods are less effective at reducing environmental impact in prejudiced populations. Our model generates thresholds for when reducing costs or emissions is more useful depending on the factors which influence the population’s opinion formation. Overall, our model provides a framework for investigating the importance of cognitive and social structures in determining human-environment dynamics.
\end{abstract}



\end{frontmatter}



\section{Introduction}
\label{sec1}

    Environmental protection is a critical issue for which consensus is needed for collective action. Public opinion is a significant determinant of public policy \citep{burstein2003impact}. However, there is a lack of consensus in public opinion on a variety of environmental issues \citep{CHIAPELLA2018169, funk2020concern, GUO2019323, buttermore2023international}. For example, surveys show that only 50-60\% of individuals in the United States believe that climate action is a high priority, with a slightly higher 60-70\% in Canada expressing the same view \citep{buttermore2023international}. Furthermore, environmental issues have become increasingly politicized, deepening polarization in public opinion \citep{doi:10.1177/1075547019900290, smith2024polarisation}.
  
    Polarization and disagreement in public opinion on environmental issues are governed by a variety of social, psychological, and experiential factors. For example, social identities are formed around agreeing individuals, further strengthening polarization \citep{bliuc2015public, Irene17072024, iyengar2012affect}.  The integration of social identity into opinion formation was shown to be the main factor in determining participation in climate strikes among youth \citep{brugger2020social}. An individual’s experiences with negative environmental issues also plays a role in their opinion formation process \citep{bergquist2019experiencing, PAPP2022102652, sloggy2021changing}. For example, experience with natural disasters, temperature anomalies and poor air quality were associated with environmental concern and voting for environmentally-motivated platforms \citep{hoffmann2022climate, PAPP2022102652}. In the United States, experiencing extreme climate events like Hurricane Irma was shown to increase belief in climate change \citep{bergquist2019experiencing}. However, individual's experiences with the environment are often shaped by their prior beliefs and prejudices \citep{CHIAPELLA2018169, denny2022severe, GUO2019323, howe2013remembers, myers2013relationship}. For example, a United States survey found that individuals who were dismissive of climate change were less inclined to report higher-than-normal temperatures during the summer \citep{howe2013remembers}. Current investigations into these factors (social influence, prejudice, and experience) have treated them independently. Further research should aim to understand the interplay of these factors in determining the environmental dynamics in polarized populations and populations in disagreement.

    Despite the importance of social and psychological factors in environmental issues, many models treat humans and the environment as uncoupled systems. Over longer time scales, such as in many environmental systems, human decision making becomes an essential factor. Recent coupled human-environment system (CHES) models aim to incorporate the effects of human decision making on the environment and vice versa. These models display a wider variety of qualitative regimes when compared to either uncoupled case \citep{tilman2020evolutionary}. CHES models have been applied towards a variety of environmental phenomena including forest conservation \citep{satake2007coupled, SIGDEL2017132}, fisheries management \citep{bailey2019computational, tilman2017maintaining}, and climate change \citep{beckage2018linking, bury2019charting, moore2022determinants}.

    One approach used in CHES models relies on evolutionary game theory \citep{bury2019charting, tilman2020evolutionary, weitz2016oscillating}. In these models, the proportion of mitigating individuals is dependent on the strategy's associated payoff. This framework has allowed for analytical approaches to study the long-term behavior of various human-environment dynamics \citep{tilman2020evolutionary}. Stability analysis of the dynamics demonstrates that long-term behavior is dependent on the incentives of mitigation and the rate of environmental feedback. To further incorporate social factors, models have factored in the benefit of complying with social norms into the net cost of mitigation \citep{bury2019charting}. 

    Opinion dynamics models provide an approach more tailored towards analyzing the opinion-formation process of individuals. Opinion dynamics models are a type of agent-based model used to study how a population’s opinions evolve over time according to social and psychological influences \citep{noorazar2020classical}. Each agent’s opinion is shaped by the opinions of the agents they interact with. The set of interacting agents is determined by a social network or by a confidence bound (meaning agents only interact with agents whose opinions are near their own). Different models incorporate additional mechanisms. In the Friedkin-Johnsen (FJ) model \citep{friedkin1990social}, each agent has a prejudice that continues to influence them during their opinion formation process. The relative weight between internal bias and social susceptability can vary across agents or populations. In the Hegselmann-Krause (HK) model with truth seekers \citep{hegselmann2006truth}, agents adjust their opinions based on a fixed ‘true’ opinion. The degree to which agents prioritize the ‘truth’ versus social influence can significantly affect how opinions evolve in the system.
    
    Despite the abundance of research on opinion dynamics models and their relevance to the factors governing public opinion on environmental issues, coupled opinion-environmental models are a recent development. A recent study coupled an adapted FJ model with climate dynamics to investigate the emergence of polarization \citep{kumar2025opinion}. They found that opinions favored mitigation after steep temperature increases. Despite an initially uniform distribution of opinions the model demonstrated that polarization pushed agent’s opinions to either extreme. 
    
    High-dimensionality makes analyzing coupled opinion-environmental models difficult. One approach to this problem has been to reduce the model to a 2-dimensional system by considering the dynamics at consensus \citep{couthures2025bifurcation, couthures2025global}. There is a need to apply this method to a more realistic setting which reflects the disagreement and polarization seen in real populations. Under this setting, we could better investigate the social, psychological, and experiential factors which lead to polarization.
    
    In this paper we develop an analytical understanding of opinion-environmental dynamics in polarized populations. We conduct stability and numerical bifurcation analysis to answer the following research questions: i) In what scenarios does polarization exist? Can polarization be reduced? ii) What is the long-term behavior of the opinion-environmental system in polarized populations? How does prejudice change this behavior? iii) What changes can be made to reduce the environmental impact in prejudiced and polarized populations? iv) In what scenarios does the population converge on the objective best response? Is it possible for a prejudiced population to converge on the objective best response?

\section{Materials and Methods}
\label{sec2}

\subsection{Model}
\label{2subsec1}
    In this paper, simplified models are used to describe opinion and environmental dynamics. This allows us to develop an analytical understanding of how the factors influencing public opinion and the environment interact. These stylized models can generate interesting effects that warrant further study. The population in our model consists of $N$ agents $A = \{1, ..., N\}$. Each agent $i \in A$ has an opinion $o_i(t) \in [-1, 1]$, we define the opinion profile as $o(t) = (o_i(t))_{i\in A}$. An agent is a mitigator if their opinion is closer to 1, and is a non-mitigator if their opinion is closer to -1.
        
    \textbf{Environmental dynamics.} We define the environmental state $e(t) \in \rm I\!R_{\geq 0}$ representing the level of a resource that naturally decays, such as a pollutant. The rate that the pollutant naturally decays is given by $\gamma$. The population's emission rate at time $t$ is $u(o(t))$. The function $u: [-1, 1]^N \rightarrow \rm I\!R_{\geq 0}$ maps the opinion profile of the population to their emission rate. We define $\tau_e$ as the relative speed of the environmental dynamics. For simplicity, we assign $\tau_e = 1$. The environmental state is governed by:
    \begin{equation}
    \tau_e\dot e(t) = -\gamma e(t) + u(o(t))
    \end{equation}
        
    \textbf{Opinion dynamics.} An agent's opinion is determined by a weighted combination of agents they interact with, personal prejudice, and the objective best response corresponding to the current environmental state. Each of these factors have been shown to influence public opinion on environmental issues in empirical studies \citep{bergquist2019experiencing, bliuc2015public, brugger2020social, CHIAPELLA2018169, GUO2019323, hoffmann2022climate, howe2013remembers, PAPP2022102652,sloggy2021changing}. Agent interaction is dependent on a confidence bound, $r$. Agents $i, j \in A$ interact if and only if the distance between their opinions is less than or equal to $r$ ($|o_i(t) - o_j(t)| \leq r$). We define the set of agents that interact with agent $i$ at time $t$ to be $N_i(t) = \{j\in A: |o_i(t) - o_j(t)| \leq r\}$. 
        
    An agent's prejudice is defined to be their initial opinion $o_i(0)$. The objective best response in our model is dynamic since it is dependent on the environmental state. The objective best response is defined by a function $h: \rm I\!R_{\geq 0} \rightarrow \{-1, 0, 1\}$ that maps the environmental state to the opinion with the highest payoff. Combining these factors, we model the opinion of each agent by:
    \begin{equation}
    \tau_o \dot{o_i(t)} = -o_i(t) + \frac{\lambda(1 - \omega)}{|N_i(t)|}\sum_{j \in N_i(t)}o_j(t) + \lambda\omega o_i(0) \ + (1-\lambda)h(e(t))
    \end{equation}
        
    The relative strength of social susceptibility, prejudice, and objectivity in opinion formation are given by the influence coefficients $\lambda(1-\omega)$, $\lambda\omega$, and $1-\lambda$ respectively. The influence coefficients lie on the simplex $\triangle^2$ since $\lambda \in [0,1]$ and $\omega \in [0,1]$. $\tau_o$ represents the speed of opinion dynamics. Similar to previous studies \citep{bury2019charting, kumar2025opinion} we will assume that the opinion dynamics evolve slower than the environmental dynamics, $\tau_o > \tau_e = 1$. For all analysis and simulations discrete versions of Eq (1) and (2) were used.

\subsection{Emission and response functions}
\label{2subsec2}

    For the results presented in Sections \ref{3subsec2}, \ref{3subsec3}, \ref{3subsec4}, \ref{3subsec5}  it is necessary to define appropriate functions $h$ and $u$. We assume that a population full of non-mitigators ($o_i(t) = -1 \ \ \forall i \in A$) emit the pollutant at a rate $\alpha$ and a population of mitigators ($o_i(t) = 1 \ \ \forall i \in A$) does not emit. Assuming that the emission rate is linearly dependent on the opinion profile we define $u(o(t)) = \frac{1}{2}\alpha(1-\frac{1}{N}\sum_{i=1}^No_i(t))$. The best response function $h$ equals 1 or -1 depending on whether the mitigator or non-mitigator opinion has a higher payoff, respectively. For simplicity, we define the payoff of switching from a non-mitigator to a mitigator opinion to be $-\beta + e(t)$ where $\beta$ is the net cost of mitigating and $e(t)$ is the level of pollutant, which harms an individual's health or psychology. Therefore, $h(e(t)) = sign(-\beta + e(t))$. 

\subsection{Visualization and numerical bifurcation analysis}
\label{2subsec3}

    Heat maps were used to visualize time-averaged values for $e(t)$ and $o(t)$ over the cost-emissions ($\beta$, $\alpha$) parameter space. The parameter space was divided into a $100 \times 100$ grid spanning $\beta \in [-10, 10]$ and $\alpha \in [0, 2]$. These values were chosen to adequately sample all regions with qualitatively distinct long-term behavior. The model was simulated for $T=1000$ steps and the first $900$ time steps were removed to eliminate transients. The remaining values for $e(t)$ and $o(t)$ were averaged and used for visualization. All simulations were initialized with $a(0) = 1, b(0)=-1, \gamma = 0.2$.
    
    Similarly, heat maps were used to visualize time-averaged values for $e(t)$ and $o(t)$ over the influence coefficients simplex. Parameter values were probed from a $100 \times 100$ grid spanning $\lambda \in [0,1]$ and $\omega \in [0,1]$, each associated with a unique triplet ($\lambda\omega$, $\lambda(1-\omega)$, $1-\lambda$) in the simplex $\triangle^2$. As before, the model was simulated for $T=1000$ steps and the last $100$ were used to obtain average values for $e(t)$ and $o(t)$. All simulations were initialized with $a(0) = 1, b(0)=-1, \gamma = 0.2$.

    Heat maps were used to visualize time-averaged values for polarization $P(t)$ over the prejudice-confidence bound ($\lambda\omega$, $r$) parameter space. The parameter space was divided into a $300 \times 300$ grid spanning $\lambda\omega \in [0, 0.5]$ and $r \in [0, 2]$. As before, the polarization map was simulated for $T=1000$ steps and the last $100$ were used to obtain average values for $P(t)$. All simulations were initialized with $a(0) = 1$, $b(0)=-1$, $\gamma = 0.2$, $1-\lambda(1-\omega) = 0.5$.
    
    Numerical bifurcation analysis was used to investigate dependence of the behavior of $e(t)$ and $o(t)$ on the relative cost of mitigation ($\theta$), the net cost of mitigation ($\beta$), and the emission rate ($\alpha$). In each case, $10000$ values were sampled from $\theta \in [0, \pi]$, $\beta \in [-10, 10]$ and $\alpha \in [0,2]$, respectively. Since $\theta$ represents an angle, a magnitude of $5$ was chosen to specify a unique point in polar coordinates in the ($\beta$, $\alpha$) parameter space. The model was simulated for $T=1000$ time steps and the first $900$ time steps were removed to eliminate transients. The remaining values are visualized in the resulting numerical bifurcation diagram.

    All simulations and visualizations were created in Python 3.13.5 using the NumPy, Matplotlib, and Mpltern packages. For further details on the numerical analysis and code see: https://github.com/CameronKerr/PolDynamics.

\subsection{Reduction of dynamics in polarized populations}
\label{2subsec4}
     We focus our analysis of the opinion-environmental dynamics on the case in which the population is polarized. In a polarized population the system presented in Eq (1) and (2) reduces to three dimensions. 
    
    \textbf{Lemma 1:} For $i, j \in A$, if $o_i(0) = o_j(0)$ then $o_i(t) = o_j(t) \ \forall t$.
    
    \textit{Proof:} We proceed by induction, the base case (t = 0) is given. Assume that for some $t \geq 0, o_i(t) = o_j(t)$. Therefore, for any $k \in A$, $|o_i(t) - o_k(t)| \leq r$ iff $|o_j(t) - o_k(t)| \leq r$. This implies $N_i(t) = N_j(t)$. Therefore:
    \begin{align*}
    o_i(t+1) &= (1 - \frac{1}{\tau_o})o_i(t) + \frac{1}{\tau_o}(\frac{\lambda(1 - \omega)}{|N_i(t)|}\sum_{k \in N_i(t)}o_k(t) + \lambda\omega o_i(0) \ + (1-\lambda)h(e(t))) \\
    &= (1 - \frac{1}{\tau_o})o_j(t) + \frac{1}{\tau_o}(\frac{\lambda(1 - \omega)}{|N_j(t)|}\sum_{k \in N_j(t)}o_k(t) + \lambda\omega o_j(0) \ + (1-\lambda)h(e(t))) \\
    &= o_j(t+1) \ ^\blacksquare
    \end{align*}
    
    Similarly to the synchronization manifold presented in \citep{couthures2025bifurcation} we introduce the polarization manifold, $L$. The polarization manifold consists of all opinion profiles and environmental states in which the opinion profile can be split into two groups where each group consists of agents with the same opinion.
    \begin{equation*}
    L = \{(o, e) \in [-1, 1]^N \times \rm I\!R_{\geq 0}\ |\ o = (a\textbf{1}_n, b\textbf{1}_{N-n}),\ a,b\in[-1,1], n \leq N \}
    \end{equation*}
    
    \textbf{Proposition 1:} If $(o(0), e(0)) \in L$, then $(o(t), e(t)) \in L \ \forall t$. 
    
    \textit{Proof:} We proceed by induction, the base case (t=0) is given. Since $(o(0), e(0)) \in L$ there exists $n$ such that $o_i(0) = o_j(0) \  \forall i,j \in \{1, ..., n\}$ and $o_i(0) = o_j(0) \ \forall i,j \in \{n+1, ..., N\}$. By Lemma 1, $o_i(t) = o_j(t) \  \forall i,j \in \{1, ..., n\}$ and $o_i(t) = o_j(t) \ \forall i,j \in \{n+1, ..., N\}$ for all $t$. Therefore, $\forall \ t$ there exists $a(t), b(t) \in [-1, 1]$ such that $o(t) = (a(t)\textbf{1}_n, b\textbf(t){1}_{N-n})$. $^\blacksquare$
    
    We can redefine the dimension $N+1$ system defined in Eq (1) and (2) to a 3-dimensional system in terms of the opinion of the two polarized groups ($a(t), b(t)$) and the environmental state ($e(t)$). Note that the dynamics change depending on whether the polarized groups are within the bound of confidence $r$. If $|a(t) - b(t)| > r$:
    \begin{align*}
    a(t+1) &= (1 - \frac{1}{\tau_o})a(t) + \frac{1}{\tau_o}(\lambda(1-\omega)a(t) + \lambda\omega a(0) + (1 - \lambda)h(e(t))) \\
    b(t+1) &= (1 - \frac{1}{\tau_o})b(t) + \frac{1}{\tau_o}(\lambda(1-\omega) b(t) + \lambda\omega b(0)+ (1 - \lambda)h(e(t))) \\
    e(t+1) &= (1 - \gamma)e(t) + u([a(t)\mathbf{1}_{n}, b(t)\mathbf{1}_{N-n}]^T)
    \end{align*}
    If $|a(t) - b(t)| \leq r$:
    \begin{align*}
    a(t+1) &= (1 - \frac{1}{\tau_o})a(t) + \frac{1}{\tau_o}(\lambda(1-\omega) (pa(t) + (1-p)b(t)) + \lambda\omega a(0) + (1 - \lambda)h(e(t))) \\
    b(t+1) &= (1 - \frac{1}{\tau_o})b(t) + \frac{1}{\tau_o}(\lambda(1-\omega) (pa(t) + (1-p)b(t)) + \lambda\omega b(0) + (1 - \lambda)h(e(t))) \\
    e(t+1) &= (1 - \gamma)e(t) + u([a(t)\mathbf{1}_{n}, b(t)\mathbf{1}_{N-n}]^T)
    \end{align*}
    
    where p is the proportion of agents in group $a(t)$, $p = \frac{n}{N}$.

\section{Results}
\label{sec3}

\subsection{Long-term existence of polarization}
\label{3subsec1}

    In this section, we show that polarization persists in a population whenever prejudice is present. Furthermore, the level of polarization at equilibrium is proportional to the strength of prejudice in the population. To reduce polarization, one must either reduce the level of prejudice or broaden the range of tolerance in which agents in each polarized group consider the opposing opinion.
    
    To formalize the notion of polarization, we define the polarization function as the distance between the two polarized opinion groups:
    \begin{equation*}
    P(t) = |a(t) - b(t)|
    \end{equation*}
    Without loss of generality, we assume $a(0) > b(0)$. From this assumption it follows that $a(t) \geq b(t)$ for all t (see Appendix A.1), allowing us to simplify the polarization function to $P(t) = a(t) - b(t)$. By substituting the expressions for $a(t)$ and $b(t)$ we can determine the difference equations governing polarization. It is important to note that any environmental influence is symmetric across both groups and therefore cancels out in the calculation of polarization.
    
    When $P(t) > r$:
    \begin{equation*}
    P(t+1) = \frac{\tau_o - 1 + \lambda(1-\omega)}{\tau_o}P(t) + \frac{\lambda\omega}{\tau_o} P(0)
    \end{equation*}
    
    When $P(t) \leq r$:
    \begin{equation*}
    P(t+1) = \frac{\tau_o - 1}{\tau_o}P(t) + \frac{\lambda\omega}{\tau_o}P(0) 
    \end{equation*}
    
    \textbf{Proposition 2}: If $\frac{\lambda\omega P(0)}{1 - \lambda(1-\omega)} <r$ then $\lim_{t \xrightarrow{} \infty}P(t) = \lambda\omega P(0)$. If $\frac{\lambda\omega P(0)}{1 - \lambda(1-\omega)} >r$ then $\lim_{t \xrightarrow{} \infty}P(t) = \frac{\lambda\omega P(0)}{1 - \lambda(1-\omega)}$. \textit{Proof}: See Appendix A.2

    This proposition, along with the heatmap of long-term polarization (Figure 1), characterizes the long-term behavior of polarization in terms of the factors influencing opinion formation. Polarization converges to 0 if and only if the population is non-prejudiced ($\lambda\omega = 0$) and has a non-zero objectivity ($1-\lambda > 0$). Conversely, if the population is prejudiced ($\lambda\omega > 0$), polarization will persist indefinitely ($\lim_{t \rightarrow \infty}P(t) > 0$). However, polarization will always decrease from its initial value $P(0)$ as long as the population exhibits some degree of objectivity. 

    \begin{figure}[H]
    \centering
    \includegraphics[width=1\linewidth]{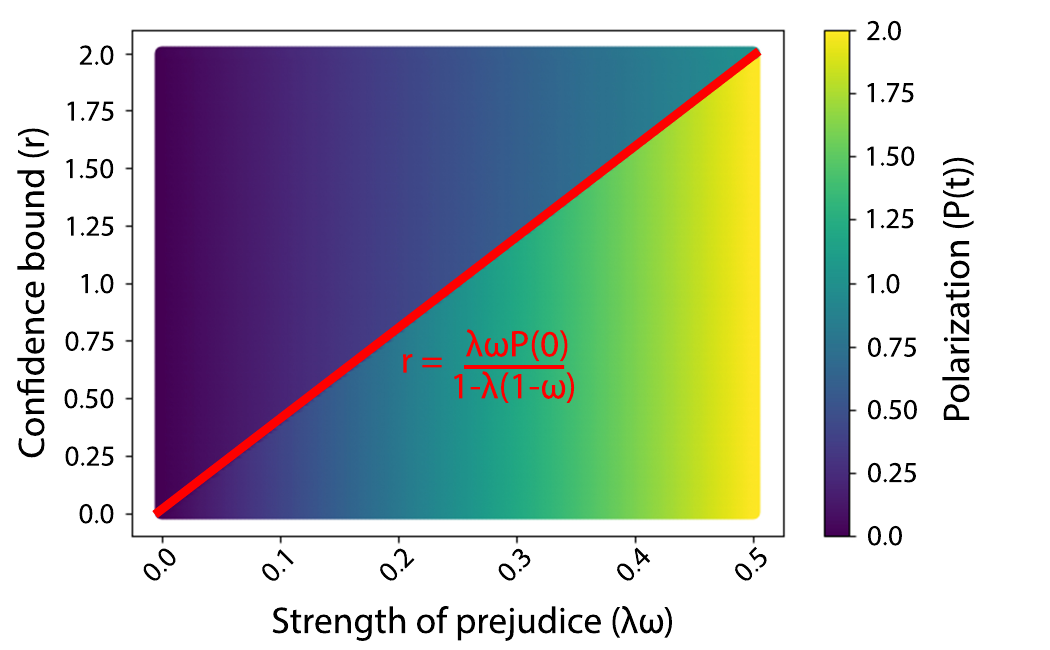}
    \caption{\textbf{Long-term behavior of polarization.} Long-term behavior of the polarization $P(t) = |a(t) - b(t)|$ between the opinion groups over the strength of prejudice ($\lambda\omega$) and the confidence bound ($r$). The sharp line represents the theoretical threshold ($r = \frac{\lambda\omega}{1-\lambda(1-\omega)}$) separating the regions of high and low polarization described in Proposition 2.}
    \label{fig:Figure1}
    \end{figure}
    
    In prejudiced populations, polarization can be reduced by increasing the confidence bound $r$ above the threshold $\frac{\lambda\omega P(0)}{1-\lambda(1-\omega)}$. Above this threshold, the two opinion groups consider each other's opinions at equilibrium. Doing so reduces polarization by a factor of $\frac{1}{1-\lambda(1-\omega)}$. In the non-prejudiced case, polarization always converges to zero, although increasing $r$ increases the rate of convergence (for details see Appendix A.3).

\subsection{Behavior of the opinion-environmental system in non-prejudiced populations}
\label{3subsec2}

  In this section, we show that the opinion profile and level of pollutant in non-prejudiced populations hinges on whether the cost of mitigation is higher or lower than the maximum environmental cost.
    
    To begin, we analyze the dynamics in non-prejudiced populations ($\lambda\omega = 0$). According to Proposition 2, the polarized groups in a non-prejudiced population converge to a consensus. Therefore, we can reduce our analysis by considering a population at consensus, let the consensus opinion be $c(t)$. Since there is no prejudice, agents value the opinion of others with a strength $\lambda \in [0,1]$ and the objective best response with strength $1-\lambda$. Incorporating our functions $h$ and $u$, our model reduces to:
    \begin{align*}
    c(t+1) &= \frac{\tau_o - 1 + \lambda}{\tau_o}c(t) + \frac{1 - \lambda}{\tau_o}sign(-\beta+e(t)) \\
    e(t+1) &= (1 - \gamma)e(t) + \frac{1}{2}\alpha(1-c(t))
    \end{align*}
    \textbf{Proposition 3}: The point $(c, e) = (1, 0)$ is a stable fixed point when $\beta < 0$ and $(c, e) = (-1, \frac{\alpha}{\gamma})$ is a stable fixed point when $\beta > \frac{\alpha}{\gamma}$. \textit{Proof}: See Appendix B.1.
    
    This proposition describes the long-term behavior in two regions of the cost-emissions ($\beta$, $\alpha$) parameter space. In the first region, there is a net benefit to mitigation ($\beta < 0$) making it the objective best response ($h(e(t)) = 1$). In this region, the proposition implies that the population holds a purely mitigative opinion ($c(t) = 1$). In the second region, where the cost of mitigating exceeds the maximum environmental cost ($\beta > \frac{\alpha}{\gamma}$), not mitigating is the objective best response ($h(e(t)) = -1$). Here, the population holds a purely non-mitigative opinion ($c(t) = -1$). We refer to the region where mitigation is the best response as the mitigative region, and the region where non-mitigation is the best response as the non-mitigative region.

    Figure 2a presents a numerical heatmap that confirms these analytical predictions. It shows sharp boundaries defining the mitigative and non-mitigative regions, consistent with the theoretical thresholds. The opinion profile and environmental state observed in each region are consistent with the equilibrium described in Proposition 3.

    \begin{figure}[H]
    \centering
    \includegraphics[width=1\linewidth]{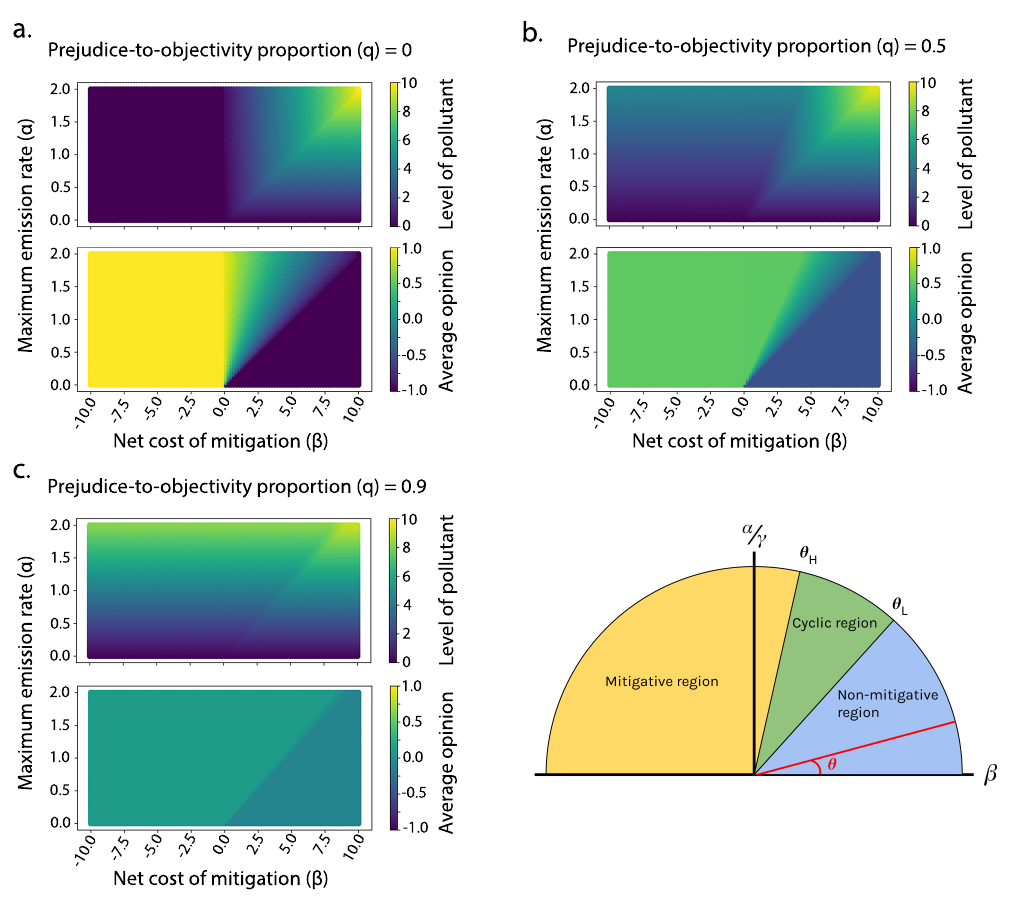}
    \caption{\textbf{Long-term behavior over the cost-emissions parameter space.} Effect of the net cost of mitigation ($\beta$) and the maximum emission rate ($\alpha$) on the level of pollutant and average opinion in the population. Heat maps visualize the long-term behavior in populations with (a) no prejudice ($q = 0$), (b) equal prejudice and objectivity ($q = 0.5$), and (c) high levels of prejudice ($q=0.9$). In each case, the behavior is separated into three regions (mitigative, cyclic, and non-mitigative regions) dependent on the angle to the $\beta$-axis, $\theta = tan^{-1}(\frac{\alpha}{\beta\gamma})$. This behavior is constant along lines from the origin in the $(\beta, \alpha)$ parameter space shown in (a-c).}
    \label{fig:Figure2}
\end{figure}
    
    We can define the region in between these cases the cyclic region. In the non-prejudiced case, this region occurs when $0 < \beta < \frac{\alpha}{\gamma}$. Here, there is a cost to mitigate, but it is 'worth it' because the potential environmental damage outweighs the cost. Numerical bifurcation diagrams (Figure 3a) show that the model exhibits cycles in this region. Specifically, the environmental state $e(t)$ oscillates above and below the cost of mitigation $\beta$ and the opinions fluctuate around the value $c(t) = 1 - \beta\frac{2\gamma}{\alpha}$. The numerical heatmap in Figure 2a supports these predictions by confirming the presence of the cyclic region in the corresponding parameter space.
    
    Proposition 3 and the numerical heat map in Figure 2a demonstrate that the mitigative, non-mitigative, and cyclic regions of the ($\beta, \alpha$) parameter space are defined by their angle to the $\beta$-axis. Let us denote this angle by $\theta$, defined as $\theta = tan^{-1}(\frac{\alpha}{\beta\gamma})$. Intuitively, $\theta$ represents how the net cost of switching to a mitigator strategy ($\beta$) compares to the maximum environmental cost ($\frac{\alpha}{\gamma}$). A low value indicates that mitigation is a costly strategy relative to the environmental damage, while a high value suggests the opposite. Therefore, we refer to $\theta$ as the \textit{relative cost of mitigation}.
    
    Our previous results can be restated in terms of the relative cost of mitigation. The non-mitigative region occurs in non-prejudiced populations when $0 < \theta < \frac{\pi}{4}$, the cyclic region when $\frac{\pi}{4} < \theta < \frac{\pi}{2}$, and the mitigative region when $\frac{\pi}{2} < \theta < \pi$. The numerical bifurcation diagram over $\theta$ shown in Figure 3a confirms these boundaries. 

\subsection{Behavior of the opinion-environmental system in prejudiced populations}
\label{3subsec3}

    In this section, we find that in a prejudiced population, each individual's opinion converges to a weighted average between their own prejudice and the objective best response. 
    
    To analyze the qualitative behavior of the system in prejudiced populations, we begin by computing numerical bifurcation diagrams with respect to the relative cost of mitigation $\theta$. These diagrams (Figure 3b and 3c) reveal that while the overall structure of the mitigative, non-mitigative, and cyclic regions is preserved from the non-prejudiced case, the boundaries and system behavior differ. Most significantly, the level of pollutant no longer reduces to zero in the mitigative region, and the polarized groups never converge to a consensus.

\begin{figure}[H]
    \centering
    \includegraphics[width=1\linewidth]{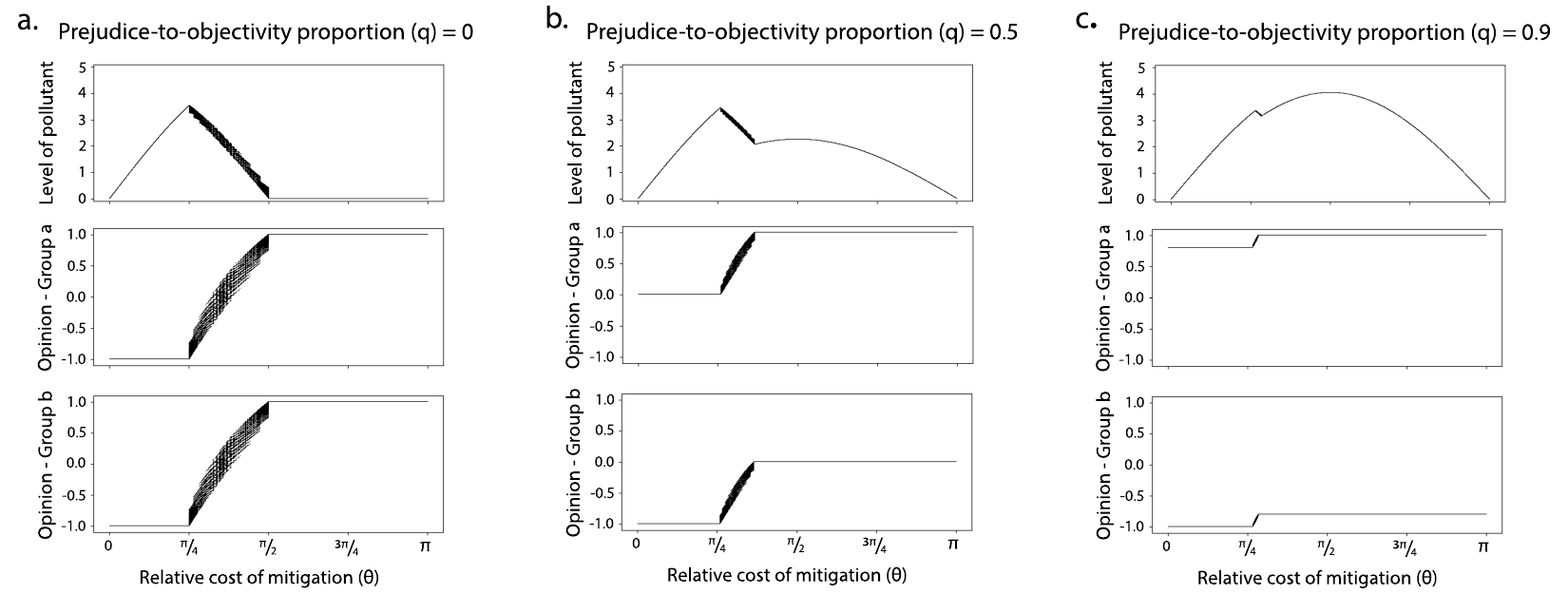}
    \caption{\textbf{Long-term behavior of the opinion-environmental system over the relative cost of mitigation ($\theta$).} Bifurcation diagrams over ($\theta$) visualize the fixed points and behaviors of the level of pollutant and opinion profile in the mitigative, non-mitigative, and cyclic regions in populations with (a) no prejudice ($q = 0$), (b) equal prejudice and objectivity ($q = 0.5$), and (c) high levels of prejudice ($q=0.9$).}
    \label{Figure3}
\end{figure}
    
    To better understand the origin of these changes, we derived analytical expressions for the fixed points in each region. Full expressions and derivations are provided in Appendix B.2. These expressions match the dynamics observed in the numerical bifurcation diagrams, as shown in Figure D.1.
    
    The analytical expressions for the opinion profile at equilibrium show that an agent's opinion converges to a weighted average of their prejudice and the objective best response. The weight assigned to an agents prejudice is given by the \textit{prejudice-to-objectivity proportion}, denoted $q = \frac{\lambda\omega}{\lambda\omega + (1-\lambda)}$. When $q > 0.5$, the population is more prejudiced than objective and when $q < 0.5$ the population is more objective than prejudiced. The objective best response is weighted by $1-q = \frac{1-\lambda}{\lambda\omega + (1-\lambda)}$. This structure leads agents in both polarized groups to adopt more mitigative opinions in the mitigative region than in the non-mitigative region, as shown in the lower panels of Figures 3b and 3c. The shift toward the objective best response is more pronounced in Figure 3b, where the prejudice-to-objectivity proportion is lower. Additionally, each polarized group remains closer to their prejudices (with $a(0) = 1, b(0) = -1$ in Figures 2 and 3) when the prejudice-to-objectivity proportion is higher, as illustrated in Figure 3c.
    
    The analytical expression for the environmental fixed point shows that the level of pollutant at equilibrium depends linearly on the prejudice-to-objectivity proportion $q$. In the mitigative region, the level of pollutant increases with $q$ (as the population becomes more prejudiced and less objective) and it decreases with $q$ in the non-mitigative region. Notably, the population's social susceptibility, given by $\lambda(1-\omega)$, has no effect on the long-term environmental outcome. We confirm this numerically by plotting a heatmap of the limiting environmental state over the influence coefficients simplex (Figure D.2). This heatmap demonstrates that the environmental outcome remains constant along lines originating from the social susceptibility vertex, along which $q$ remains constant. 
    
    In the cyclic region, cycles still occur in prejudiced populations (Figure 3b, 3c) but their structure is altered. The environmental state continues to oscillate around the cost of mitigation $e(t) = \beta$ and the average opinion oscillates around the same mean as in the non-prejudiced case, $c(t) = 1-\beta\frac{2\gamma}{\alpha}$. However, unlike the non-prejudiced case, the two polarized groups do not converge to this average. Instead, each group is symmetrically pulled toward its prejudice.

\subsection{Pathways to lower environmental impact in polarized and prejudiced populations}
\label{3subsec4}

    In this section, we show that reducing the level of pollutant by lowering the cost of mitigation is less effective in populations that are prejudiced and polarized. 
    
    The fixed-point expressions for the environment reveal thresholds that determine whether changes in the emission rate ($\alpha$) or cost of mitigation ($\beta$) reduce the level of pollutant. Increasing the emission rate raises the level of pollutant when ($\beta$, $\alpha$) are in the non-mitigative or mitigative region and has no impact in the cyclic region. Conversely, increasing the cost of mitigation increases the level of pollutant when ($\beta$, $\alpha$) are in the cyclic region and has no impact in the mitigative or non-mitigative region. These regions of sensitivity and stagnation are visualized in the numerical bifurcation diagrams of $\beta$ and $\alpha$ (Figure 4). Overall, the effect of a parameter change is context-dependent. For instance, if the cost of mitigation is higher than the maximum environmental damage ($\beta > \frac{\alpha}{\gamma}$) then reducing the cost has no effect on level of pollutant unless it passes a threshold that shifts the system into the cyclic region.

\begin{figure}[H]
    \centering
    \includegraphics[width=1\linewidth]{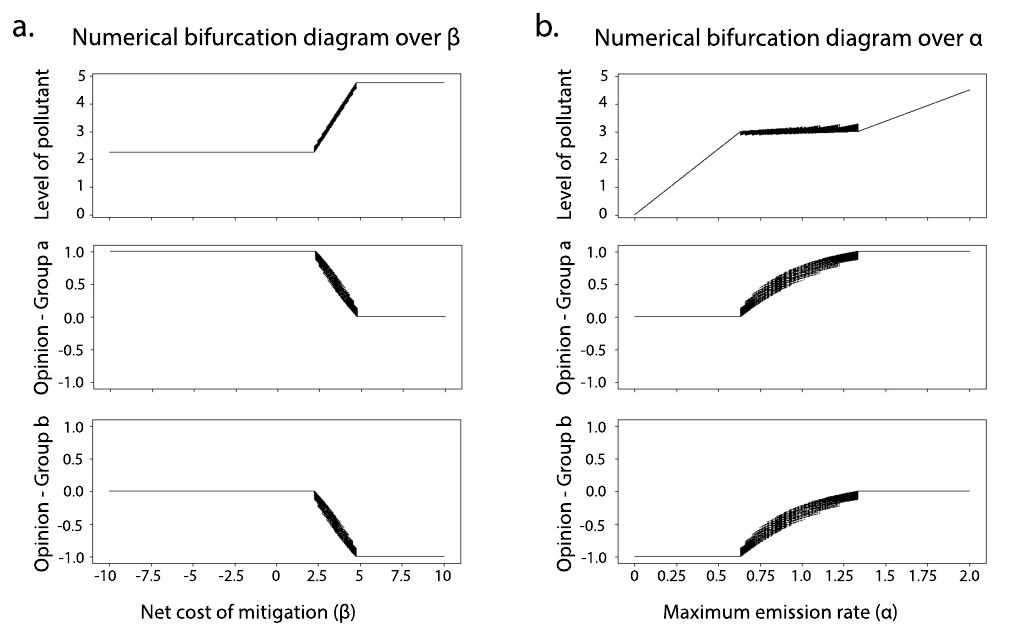}
    \caption{\textbf{Regions of sensitivity and stagnation in the opinion-environmental system over the cost-emissions parameter space.} Bifurcation diagrams visualize fixed points and behaviors of the level of pollutant and opinion profile over different values of the net cost of mitigation ($\beta$) and the maximum emission rate ($\alpha$).}
    \label{Figure4}
\end{figure}
    
    These thresholds correspond to the boundaries of the cyclic region, which lie between the mitigative and non-mitigative regions (defined by $h(e(t)) = 1$ and $h(e(t)) = -1$, respectively). Using these definitions, the lower and upper boundaries of the cyclic region are given by:
    \begin{align*}
    \theta_L &= tan^{-1}(\frac{2}{2-q(1+pa(0)+(1-p)b(0))}) \\ \theta_H &= tan^{-1}(\frac{2}{q(1-pa(0)-(1-p)b(0))})
    \end{align*}
    These expressions show that as the prejudice-to-objectivity proportion ($q$) increases, the lower boundary $\theta_L$ increases and the upper boundary $\theta_H$ decreases. As a result, the cyclic region narrows, which is visible in Figures 2 and 3.
    
    Since only the cyclic region is sensitive to changes in the cost of mitigation, we conclude that as prejudice increases relative to objectivity, lowering the cost of mitigation has less of an effect on the level of pollutant. This is confirmed in the heatmaps in Figure 2. For low values of the prejudice-to-objectivity proportion, the level of pollutant varies along both the cost and emissions axes (Figure 2a). As prejudice increases, the heatmap shows a vertical gradient, with contours aligned horizontally (Figure 2c). This indicates that the level of pollutant becomes invariant with respect to the cost of mitigation and depend only on the emission rate. 

\subsection{Convergence to the objective best-response}
\label{3subsec5}

    In this section, we show that the more prejudiced a population is, the further their opinion profile is from the objective best response. In the absence of prejudice, the population converges to the best response.
    
    To quantify the deviation of an opinion profile $o(t)$ from the objective best response $T \in [-1, 1]$ we define the distance function $d_T(o(t)) = \frac{1}{N}\sum_{i=1}^N|o_i(t) - T|$. In a polarized population, the function reduces to $d_T(o(t)) = p|a(t) - T| + (1-p)|b(t) - T|$. Note that the best response $T = h(e(t))$ is only well-defined in the mitigative and non-mitigative regions, where it remains constant at 1 or -1, respectively. In the cyclic region, the environmental state oscillates above and below the cost of mitigation, causing the best response to switch between favoring mitigation and non-mitigation. Therefore, we focus our analysis on the mitigative and non-mitigative regions.
    
    By substituting in the fixed-point expressions from Section \ref{3subsec3}, we derive analytical formulas for the distance to the objective best response in each region (for more details see Appendix C). In both the mitigative and non-mitigative regions, the distance grows linearly with the prejudice-to-objectivity proportion ($q$). More specifically, when the polarized groups are farther than the confidence bound at equilibrium (i.e $\frac{\lambda\omega P(0)}{1-\lambda(1-\omega)} > r$) the distance from the objective best response increases linearly with $q$. When the groups fall within each other's confidence bound (i.e $\frac{\lambda\omega P(0)}{1-\lambda(1-\omega)} \leq r$) the same linear relationship with $q$ holds, provided that the social susceptibility ($\lambda(1-\omega)$ remains fixed. In both cases, a polarized population converges to the objective best response if and only if the prejudice-to-objectivity proportion is 0 ($q = 0$).

\section{Discussion}
\label{sec4}

In this study, we developed an analytical framework for understanding opinion-environmental dynamics, focusing on the impact of social susceptibility, prejudice, and objectivity. We showed that polarization persists as long as individuals remain prejudiced. We then characterized the long-term behaviour of the opinion-environmental system, finding that the outcomes depend on the cost of mitigation versus environmental damage. Additionally, individual opinions converge to a weighted average of their personal prejudice and the objective best response. We also show that reducing the cost of mitigation becomes less effective as a tool to reduce the level of pollutant in prejudiced and polarized populations, and that these populations are further from the objective best response. Together, these results highlight how important the factors governing opinion formation are in environmental dynamics.

One of the key results from our study is the strong link between prejudice and persistent polarization. In the absence of prejudice, the individuals in the population converge to a consensus, even when objectivity is arbitrarily small. By contrast, the presence of even minimal prejudice in a population results in indefinite polarization, no matter how objective individuals are. Furthermore, prejudice distances an individual’s opinion from the objective best response, no matter what their prejudiced opinion is. These points suggest that lowering the role of personal prejudices in opinion formation is crucial for consensus on the optimal strategy.

We found that incentivizing mitigation has less of an impact on the environment in prejudiced and polarized populations. This finding adds nuance to previous results which emphasize cost reduction as a pathway to environmental action \citep{bury2019charting, kumar2025opinion}. In our model, agents who rely more on their prejudice than the objective best response are less dependent on the associated cost of their opinion. As a result, policies aimed at reducing environmental impact by lowering costs may have difficulties if the opinions of individuals are strongly prejudiced. More broadly, our analysis shows that the effectiveness of reducing costs or the maximum emissions rate is dependent on  the current costs and emissions as well as the social and psychological factors underlying opinion formation.

Future extensions of opinion-environmental dynamics could incorporate additional mechanisms to capture more realistic opinion-environment dynamics and explore new research areas. For instance, to study how subjective perception of the environment affects opinion formation, the cost function could depend on the perceived, rather than actual, environmental state. To examine the influence of media or political actors, the prejudice term could be used to represent the fixed opinion of a stubborn agent representing a news source or political party. Further simulation-based approaches would allow for more flexibility in the population, and a relaxing of this study’s simplifying assumptions. Heterogeneity in opinion formation could be incorporated by varying the strength of social-susceptibility, prejudice, and objectivity by individual. Furthermore, initial opinions could be drawn from a bimodal distribution rather than assuming strictly polarized groups.

\section{Funding Sources}
This work was supported by the Natural Sciences and Engineering Research Council (USRA - 604233 - 2025 and NSERC Discovery Grant to C.T. Bauch. RGPIN-2025-04274)
) and the Ontario Graduate Scholarship. 

\bibliographystyle{elsarticle-harv} 
\bibliography{bibliography.bib}

\begin{thebibliography}{33}
\expandafter\ifx\csname natexlab\endcsname\relax\def\natexlab#1{#1}\fi
\providecommand{\url}[1]{\texttt{#1}}
\providecommand{\href}[2]{#2}
\providecommand{\path}[1]{#1}
\providecommand{\DOIprefix}{doi:}
\providecommand{\ArXivprefix}{arXiv:}
\providecommand{\URLprefix}{URL: }
\providecommand{\Pubmedprefix}{pmid:}
\providecommand{\doi}[1]{\href{http://dx.doi.org/#1}{\path{#1}}}
\providecommand{\Pubmed}[1]{\href{pmid:#1}{\path{#1}}}
\providecommand{\bibinfo}[2]{#2}
\ifx\xfnm\relax \def\xfnm[#1]{\unskip,\space#1}\fi
\bibitem[{Bailey et~al.(2019)Bailey, Carrella, Axtell, Burgess, Cabral, Drexler, Dorsett, Madsen, Merkl and Saul}]{bailey2019computational}
\bibinfo{author}{Bailey, R.M.}, \bibinfo{author}{Carrella, E.}, \bibinfo{author}{Axtell, R.}, \bibinfo{author}{Burgess, M.G.}, \bibinfo{author}{Cabral, R.B.}, \bibinfo{author}{Drexler, M.}, \bibinfo{author}{Dorsett, C.}, \bibinfo{author}{Madsen, J.K.}, \bibinfo{author}{Merkl, A.}, \bibinfo{author}{Saul, S.}, \bibinfo{year}{2019}.
\newblock \bibinfo{title}{A computational approach to managing coupled human--environmental systems: the poseidon model of ocean fisheries}.
\newblock \bibinfo{journal}{Sustainability Science} \bibinfo{volume}{14}, \bibinfo{pages}{259--275}.
\newblock \DOIprefix\doi{https://doi.org/10.1007/s11625-018-0579-9}.
\bibitem[{Beckage et~al.(2018)Beckage, Gross, Lacasse, Carr, Metcalf, Winter, Howe, Fefferman, Franck, Zia, Kinzig and Hoffman}]{beckage2018linking}
\bibinfo{author}{Beckage, B.}, \bibinfo{author}{Gross, L.J.}, \bibinfo{author}{Lacasse, K.}, \bibinfo{author}{Carr, E.}, \bibinfo{author}{Metcalf, S.S.}, \bibinfo{author}{Winter, J.M.}, \bibinfo{author}{Howe, P.D.}, \bibinfo{author}{Fefferman, N.}, \bibinfo{author}{Franck, T.}, \bibinfo{author}{Zia, A.}, \bibinfo{author}{Kinzig, A.}, \bibinfo{author}{Hoffman, F.M.}, \bibinfo{year}{2018}.
\newblock \bibinfo{title}{Linking models of human behaviour and climate alters projected climate change}.
\newblock \bibinfo{journal}{Nature Climate Change} \bibinfo{volume}{8}, \bibinfo{pages}{79--84}.
\newblock \DOIprefix\doi{https://doi.org/10.1038/s41558-017-0031-7}.
\bibitem[{Bergquist et~al.(2019)Bergquist, Nilsson and Schultz}]{bergquist2019experiencing}
\bibinfo{author}{Bergquist, M.}, \bibinfo{author}{Nilsson, A.}, \bibinfo{author}{Schultz, P.W.}, \bibinfo{year}{2019}.
\newblock \bibinfo{title}{Experiencing a severe weather event increases concern about climate change}.
\newblock \bibinfo{journal}{Frontiers in psychology} \bibinfo{volume}{10}, \bibinfo{pages}{220}.
\newblock \DOIprefix\doi{https://doi.org/10.3389/fpsyg.2019.00220}.
\bibitem[{Bliuc et~al.(2015)Bliuc, McGarty, Thomas, Lala, Berndsen and Misajon}]{bliuc2015public}
\bibinfo{author}{Bliuc, A.M.}, \bibinfo{author}{McGarty, C.}, \bibinfo{author}{Thomas, E.F.}, \bibinfo{author}{Lala, G.}, \bibinfo{author}{Berndsen, M.}, \bibinfo{author}{Misajon, R.}, \bibinfo{year}{2015}.
\newblock \bibinfo{title}{Public division about climate change rooted in conflicting socio-political identities}.
\newblock \bibinfo{journal}{Nature Climate Change} \bibinfo{volume}{5}, \bibinfo{pages}{226--229}.
\newblock \DOIprefix\doi{https://doi.org/10.1038/nclimate2507}.
\bibitem[{Br{\"u}gger et~al.(2020)Br{\"u}gger, Gubler, Steentjes and Capstick}]{brugger2020social}
\bibinfo{author}{Br{\"u}gger, A.}, \bibinfo{author}{Gubler, M.}, \bibinfo{author}{Steentjes, K.}, \bibinfo{author}{Capstick, S.B.}, \bibinfo{year}{2020}.
\newblock \bibinfo{title}{Social identity and risk perception explain participation in the swiss youth climate strikes}.
\newblock \bibinfo{journal}{Sustainability} \bibinfo{volume}{12}, \bibinfo{pages}{10605}.
\newblock \DOIprefix\doi{https://doi.org/10.3390/su122410605}.
\bibitem[{Burstein(2003)}]{burstein2003impact}
\bibinfo{author}{Burstein, P.}, \bibinfo{year}{2003}.
\newblock \bibinfo{title}{The impact of public opinion on public policy: A review and an agenda}.
\newblock \bibinfo{journal}{Political Research Quarterly} \bibinfo{volume}{56}, \bibinfo{pages}{29--40}.
\newblock \DOIprefix\doi{https://doi.org/10.2307/3219881}.
\bibitem[{Bury et~al.(2019)Bury, Bauch and Anand}]{bury2019charting}
\bibinfo{author}{Bury, T.M.}, \bibinfo{author}{Bauch, C.T.}, \bibinfo{author}{Anand, M.}, \bibinfo{year}{2019}.
\newblock \bibinfo{title}{Charting pathways to climate change mitigation in a coupled socio-climate model}.
\newblock \bibinfo{journal}{PLoS Computational Biology} \bibinfo{volume}{15}, \bibinfo{pages}{e1007000}.
\newblock \DOIprefix\doi{https://doi.org/10.1371/journal.pcbi.1007000}.
\bibitem[{Chiapella et~al.(2018)Chiapella, Nielsen-Pincus and Strecker}]{CHIAPELLA2018169}
\bibinfo{author}{Chiapella, A.M.}, \bibinfo{author}{Nielsen-Pincus, M.}, \bibinfo{author}{Strecker, A.L.}, \bibinfo{year}{2018}.
\newblock \bibinfo{title}{Public perceptions of mountain lake fisheries management in national parks}.
\newblock \bibinfo{journal}{Journal of Environmental Management} \bibinfo{volume}{226}, \bibinfo{pages}{169--179}.
\newblock \DOIprefix\doi{https://doi.org/10.1016/j.jenvman.2018.08.040}.
\bibitem[{Chinn et~al.(2020)Chinn, Hart and Soroka}]{doi:10.1177/1075547019900290}
\bibinfo{author}{Chinn, S.}, \bibinfo{author}{Hart, P.S.}, \bibinfo{author}{Soroka, S.}, \bibinfo{year}{2020}.
\newblock \bibinfo{title}{Politicization and polarization in climate change news content, 1985-2017}.
\newblock \bibinfo{journal}{Science Communication} \bibinfo{volume}{42}, \bibinfo{pages}{112--129}.
\newblock \DOIprefix\doi{https://doi.org/10.1177/1075547019900290}.
\bibitem[{Couthures et~al.(2025a)Couthures, Bizyaeva, Varma, Franci and Morarescu}]{couthures2025bifurcation}
\bibinfo{author}{Couthures, A.}, \bibinfo{author}{Bizyaeva, A.}, \bibinfo{author}{Varma, V.S.}, \bibinfo{author}{Franci, A.}, \bibinfo{author}{Morarescu, I.C.}, \bibinfo{year}{2025}a.
\newblock \bibinfo{title}{Bifurcation analysis of an opinion dynamics model coupled with an environmental dynamics}.
\newblock \bibinfo{journal}{arXiv preprint} \DOIprefix\doi{https://doi.org/10.48550/arXiv.2504.03419}.
\bibitem[{Couthures et~al.(2025b)Couthures, Varma, Lasaulce and Morarescu}]{couthures2025global}
\bibinfo{author}{Couthures, A.}, \bibinfo{author}{Varma, V.S.}, \bibinfo{author}{Lasaulce, S.}, \bibinfo{author}{Morarescu, I.C.}, \bibinfo{year}{2025}b.
\newblock \bibinfo{title}{Global synchronization of multi-agent systems with nonlinear interactions}.
\newblock \bibinfo{journal}{IEEE Control Systems Letters} \bibinfo{volume}{9}, \bibinfo{pages}{354--359}.
\newblock \DOIprefix\doi{https://doi.org/10.1109/LCSYS.2025.3573563}.
\bibitem[{Denny et~al.(2022)Denny, Marchese and Fischer}]{denny2022severe}
\bibinfo{author}{Denny, R.C.}, \bibinfo{author}{Marchese, J.}, \bibinfo{author}{Fischer, A.P.}, \bibinfo{year}{2022}.
\newblock \bibinfo{title}{Severe weather experience and climate change belief among small woodland owners: A study of reciprocal effects}.
\newblock \bibinfo{journal}{Weather, Climate, and Society} \bibinfo{volume}{14}, \bibinfo{pages}{1065--1082}.
\newblock \DOIprefix\doi{https://doi.org/10.1175/WCAS-D-21-0176.1}.
\bibitem[{Friedkin and Johnsen(1990)}]{friedkin1990social}
\bibinfo{author}{Friedkin, N.E.}, \bibinfo{author}{Johnsen, E.C.}, \bibinfo{year}{1990}.
\newblock \bibinfo{title}{Social influence and opinions}.
\newblock \bibinfo{journal}{Journal of Mathematical Sociology} \bibinfo{volume}{15}, \bibinfo{pages}{193--206}.
\newblock \DOIprefix\doi{https://doi.org/10.1080/0022250X.1990.9990069}.
\bibitem[{Funk et~al.(2020)Funk, Tyson, Kennedy and Johnson}]{funk2020concern}
\bibinfo{author}{Funk, C.}, \bibinfo{author}{Tyson, A.}, \bibinfo{author}{Kennedy, B.}, \bibinfo{author}{Johnson, C.}, \bibinfo{year}{2020}.
\newblock \bibinfo{title}{Concern over climate and the environment predominates among these publics}.
\newblock \bibinfo{journal}{Pew Research Center} \URLprefix \url{https://www.pewresearch.org/science/2020/09/29/concern-over-climate-and-the-environment-predominates-among-these-publics/}. \bibinfo{note}{last accessed: 2025-09-23}.
\bibitem[{Guo et~al.(2019)Guo, Gill, Johengen and Cardinale}]{GUO2019323}
\bibinfo{author}{Guo, T.}, \bibinfo{author}{Gill, D.}, \bibinfo{author}{Johengen, T.H.}, \bibinfo{author}{Cardinale, B.L.}, \bibinfo{year}{2019}.
\newblock \bibinfo{title}{What determines the public’s support for water quality regulations to mitigate agricultural runoff?}
\newblock \bibinfo{journal}{Environmental Science \& Policy} \bibinfo{volume}{101}, \bibinfo{pages}{323--330}.
\newblock \DOIprefix\doi{https://doi.org/10.1016/j.envsci.2019.09.008}.
\bibitem[{Hegselmann and Krause(2006)}]{hegselmann2006truth}
\bibinfo{author}{Hegselmann, R.}, \bibinfo{author}{Krause, U.}, \bibinfo{year}{2006}.
\newblock \bibinfo{title}{Truth and cognitive division of labor: First steps towards a computer aided social epistemology}.
\newblock \bibinfo{journal}{Journal of Artificial Societies and Social Simulation} \bibinfo{volume}{9}, \bibinfo{pages}{10}.
\newblock \URLprefix \url{https://www.jasss.org/9/3/10/10.pdf}.
\bibitem[{Hoffmann et~al.(2022)Hoffmann, Muttarak, Peisker and Stanig}]{hoffmann2022climate}
\bibinfo{author}{Hoffmann, R.}, \bibinfo{author}{Muttarak, R.}, \bibinfo{author}{Peisker, J.}, \bibinfo{author}{Stanig, P.}, \bibinfo{year}{2022}.
\newblock \bibinfo{title}{Climate change experiences raise environmental concerns and promote green voting}.
\newblock \bibinfo{journal}{Nature Climate Change} \bibinfo{volume}{12}, \bibinfo{pages}{148--155}.
\newblock \DOIprefix\doi{https://doi.org/10.1038/s41558-021-01263-8}.
\bibitem[{Howe and Leiserowitz(2013)}]{howe2013remembers}
\bibinfo{author}{Howe, P.D.}, \bibinfo{author}{Leiserowitz, A.}, \bibinfo{year}{2013}.
\newblock \bibinfo{title}{Who remembers a hot summer or a cold winter? the asymmetric effect of beliefs about global warming on perceptions of local climate conditions in the us}.
\newblock \bibinfo{journal}{Global environmental change} \bibinfo{volume}{23}, \bibinfo{pages}{1488--1500}.
\newblock \DOIprefix\doi{https://doi.org/10.1016/j.gloenvcha.2013.09.014}.
\bibitem[{Irene et~al.(2024)Irene, Daniels, Irene, Kelly and Frank}]{Irene17072024}
\bibinfo{author}{Irene, J.O.}, \bibinfo{author}{Daniels, C.}, \bibinfo{author}{Irene, B.N.O.}, \bibinfo{author}{Kelly, M.}, \bibinfo{author}{Frank, R.}, \bibinfo{year}{2024}.
\newblock \bibinfo{title}{A social identity approach to understanding sustainability and environmental behaviours in south africa}.
\newblock \bibinfo{journal}{Local Environment} \bibinfo{volume}{0}, \bibinfo{pages}{1--19}.
\newblock \DOIprefix\doi{https://doi.org/10.1080/13549839.2024.2376546}.
\bibitem[{Iyengar et~al.(2012)Iyengar, Sood and Lelkes}]{iyengar2012affect}
\bibinfo{author}{Iyengar, S.}, \bibinfo{author}{Sood, G.}, \bibinfo{author}{Lelkes, Y.}, \bibinfo{year}{2012}.
\newblock \bibinfo{title}{Affect, not ideology: A social identity perspective on polarization}.
\newblock \bibinfo{journal}{Public opinion quarterly} \bibinfo{volume}{76}, \bibinfo{pages}{405--431}.
\newblock \DOIprefix\doi{https://doi.org/10.1093/poq/nfs038}.
\bibitem[{Kumar et~al.(2025)Kumar, Josi{\'c}, Bauch and Anand}]{kumar2025opinion}
\bibinfo{author}{Kumar, A.S.}, \bibinfo{author}{Josi{\'c}, K.}, \bibinfo{author}{Bauch, C.T.}, \bibinfo{author}{Anand, M.}, \bibinfo{year}{2025}.
\newblock \bibinfo{title}{From opinion polarization to climate action: A social-climate model of the opinion spectrum}.
\newblock \bibinfo{journal}{arXiv preprint} \DOIprefix\doi{https://doi.org/10.48550/arXiv.2503.04689}.
\bibitem[{Leiserowitz et~al.(2023)Leiserowitz, Verner, Wood, Carman, Ordaz~Reynoso, Thulin, Rosenthal, Marlon and Buttermore}]{buttermore2023international}
\bibinfo{author}{Leiserowitz, A.}, \bibinfo{author}{Verner, M.}, \bibinfo{author}{Wood, E.}, \bibinfo{author}{Carman, J.}, \bibinfo{author}{Ordaz~Reynoso, N.}, \bibinfo{author}{Thulin, E.}, \bibinfo{author}{Rosenthal, S.}, \bibinfo{author}{Marlon, J.}, \bibinfo{author}{Buttermore, N.}, \bibinfo{year}{2023}.
\newblock \bibinfo{title}{International public opinion on climate change, 2023}.
\newblock \bibinfo{journal}{New Haven, CT: Yale Program on Climate Change communication and Data for Good at Meta} \URLprefix \url{https://climatecommunication.yale.edu/publications/international-public-opinion-on-climate-change-2023/}. \bibinfo{note}{last accessed: 2025-09-23}.
\bibitem[{Moore et~al.(2022)Moore, Lacasse, Mach, Shin, Gross and Beckage}]{moore2022determinants}
\bibinfo{author}{Moore, F.C.}, \bibinfo{author}{Lacasse, K.}, \bibinfo{author}{Mach, K.J.}, \bibinfo{author}{Shin, Y.A.}, \bibinfo{author}{Gross, L.J.}, \bibinfo{author}{Beckage, B.}, \bibinfo{year}{2022}.
\newblock \bibinfo{title}{Determinants of emissions pathways in the coupled climate-social system}.
\newblock \bibinfo{journal}{Nature} \bibinfo{volume}{603}, \bibinfo{pages}{103--111}.
\newblock \DOIprefix\doi{https://doi.org/10.1038/s41586-022-04423-8}.
\bibitem[{Myers et~al.(2013)Myers, Maibach, Roser-Renouf, Akerlof and Leiserowitz}]{myers2013relationship}
\bibinfo{author}{Myers, T.A.}, \bibinfo{author}{Maibach, E.W.}, \bibinfo{author}{Roser-Renouf, C.}, \bibinfo{author}{Akerlof, K.}, \bibinfo{author}{Leiserowitz, A.A.}, \bibinfo{year}{2013}.
\newblock \bibinfo{title}{The relationship between personal experience and belief in the reality of global warming}.
\newblock \bibinfo{journal}{Nature climate change} \bibinfo{volume}{3}, \bibinfo{pages}{343--347}.
\newblock \DOIprefix\doi{https://doi.org/10.1038/nclimate1754}.
\bibitem[{Noorazar et~al.(2020)Noorazar, Vixie, Talebanpour and Hu}]{noorazar2020classical}
\bibinfo{author}{Noorazar, H.}, \bibinfo{author}{Vixie, K.R.}, \bibinfo{author}{Talebanpour, A.}, \bibinfo{author}{Hu, Y.}, \bibinfo{year}{2020}.
\newblock \bibinfo{title}{From classical to modern opinion dynamics}.
\newblock \bibinfo{journal}{International Journal of Modern Physics C} \bibinfo{volume}{31}, \bibinfo{pages}{2050101}.
\newblock \DOIprefix\doi{https://doi.org/10.1142/S0129183120501016}.
\bibitem[{Papp(2022)}]{PAPP2022102652}
\bibinfo{author}{Papp, Z.}, \bibinfo{year}{2022}.
\newblock \bibinfo{title}{Environmental attitudes, environmental problems and party choice. a large-n comparative study}.
\newblock \bibinfo{journal}{Political Geography} \bibinfo{volume}{97}, \bibinfo{pages}{102652}.
\newblock \DOIprefix\doi{https://doi.org/10.1016/j.polgeo.2022.102652}.
\bibitem[{Satake et~al.(2007)Satake, Leslie, Iwasa and Levin}]{satake2007coupled}
\bibinfo{author}{Satake, A.}, \bibinfo{author}{Leslie, H.M.}, \bibinfo{author}{Iwasa, Y.}, \bibinfo{author}{Levin, S.A.}, \bibinfo{year}{2007}.
\newblock \bibinfo{title}{Coupled ecological--social dynamics in a forested landscape: Spatial interactions and information flow}.
\newblock \bibinfo{journal}{Journal of Theoretical Biology} \bibinfo{volume}{246}, \bibinfo{pages}{695--707}.
\newblock \DOIprefix\doi{https://doi.org/10.1016/j.jtbi.2007.01.014}.
\bibitem[{Sigdel et~al.(2017)Sigdel, Anand and Bauch}]{SIGDEL2017132}
\bibinfo{author}{Sigdel, R.}, \bibinfo{author}{Anand, M.}, \bibinfo{author}{Bauch, C.T.}, \bibinfo{year}{2017}.
\newblock \bibinfo{title}{Competition between injunctive social norms and conservation priorities gives rise to complex dynamics in a model of forest growth and opinion dynamics}.
\newblock \bibinfo{journal}{Journal of Theoretical Biology} \bibinfo{volume}{432}, \bibinfo{pages}{132--140}.
\newblock \DOIprefix\doi{https://doi.org/10.1016/j.jtbi.2017.07.029}.
\bibitem[{Sloggy et~al.(2021)Sloggy, Suter, Rad, Manning and Goemans}]{sloggy2021changing}
\bibinfo{author}{Sloggy, M.R.}, \bibinfo{author}{Suter, J.F.}, \bibinfo{author}{Rad, M.R.}, \bibinfo{author}{Manning, D.T.}, \bibinfo{author}{Goemans, C.}, \bibinfo{year}{2021}.
\newblock \bibinfo{title}{Changing opinions on a changing climate: the effects of natural disasters on public perceptions of climate change}.
\newblock \bibinfo{journal}{Climatic Change} \bibinfo{volume}{168}, \bibinfo{pages}{25}.
\newblock \DOIprefix\doi{https://doi.org/10.1007/s10584-021-03242-6}.
\bibitem[{Smith et~al.(2024)Smith, Bognar and Mayer}]{smith2024polarisation}
\bibinfo{author}{Smith, E.K.}, \bibinfo{author}{Bognar, M.J.}, \bibinfo{author}{Mayer, A.P.}, \bibinfo{year}{2024}.
\newblock \bibinfo{title}{Polarisation of climate and environmental attitudes in the united states, 1973-2022}.
\newblock \bibinfo{journal}{NPJ Climate Action} \bibinfo{volume}{3}, \bibinfo{pages}{2}.
\newblock \DOIprefix\doi{https://doi.org/10.1038/s44168-023-00074-1}.
\bibitem[{Tilman et~al.(2020)Tilman, Plotkin and Ak{\c{c}}ay}]{tilman2020evolutionary}
\bibinfo{author}{Tilman, A.R.}, \bibinfo{author}{Plotkin, J.B.}, \bibinfo{author}{Ak{\c{c}}ay, E.}, \bibinfo{year}{2020}.
\newblock \bibinfo{title}{Evolutionary games with environmental feedbacks}.
\newblock \bibinfo{journal}{Nature Communications} \bibinfo{volume}{11}, \bibinfo{pages}{915}.
\newblock \DOIprefix\doi{https://doi.org/10.1038/s41467-020-14531-6}.
\bibitem[{Tilman et~al.(2017)Tilman, Watson and Levin}]{tilman2017maintaining}
\bibinfo{author}{Tilman, A.R.}, \bibinfo{author}{Watson, J.R.}, \bibinfo{author}{Levin, S.}, \bibinfo{year}{2017}.
\newblock \bibinfo{title}{Maintaining cooperation in social-ecological systems: effective bottom-up management often requires sub-optimal resource use}.
\newblock \bibinfo{journal}{Theoretical Ecology} \bibinfo{volume}{10}, \bibinfo{pages}{155--165}.
\newblock \DOIprefix\doi{https://doi.org/10.1007/s12080-016-0318-8}.
\bibitem[{Weitz et~al.(2016)Weitz, Eksin, Paarporn, Brown and Ratcliff}]{weitz2016oscillating}
\bibinfo{author}{Weitz, J.S.}, \bibinfo{author}{Eksin, C.}, \bibinfo{author}{Paarporn, K.}, \bibinfo{author}{Brown, S.P.}, \bibinfo{author}{Ratcliff, W.C.}, \bibinfo{year}{2016}.
\newblock \bibinfo{title}{An oscillating tragedy of the commons in replicator dynamics with game-environment feedback}.
\newblock \bibinfo{journal}{Proceedings of the National Academy of Sciences} \bibinfo{volume}{113}, \bibinfo{pages}{E7518--E7525}.
\newblock \DOIprefix\doi{https://doi.org/10.1073/pnas.1604096113}.

\end{thebibliography}


\appendix
\section{}
\label{appA}

\subsection{Proof of Lemma}
\label{appA1}
We want to prove that if $a(0) > b(0)$ then $a(t) \geq b(t)\ \forall \ t$. We will proceed by induction. The base case (t = 0) is given. Assume that for some $t\geq0$ that $a(t) \geq b(t)$.
If $|a(t) - b(t)| > r:$
\begin{align*}
&a(t+1) = \frac{\tau_o - 1 + \lambda(1-\omega)}{\tau_o}a(t) + \frac{\lambda\omega}{\tau_o}a(0) + \frac{1-\lambda}{\tau_o}h(e(t)) \\
&b(t+1) = \frac{\tau_o - 1 + \lambda(1-\omega)}{\tau_o}b(t) + \frac{\lambda\omega}{\tau_o}b(0) + \frac{1-\lambda}{\tau_o}h(e(t)) \\
&\implies a(t+1) - b(t+1) = \frac{\tau_o - 1 + \lambda(1-\omega)}{\tau_o}(a(t) - b(t)) + \frac{\lambda\omega}{\tau_o}(a(0) - b(0)) \geq 0 \\
&\implies a(t+1) \geq b(t+1)
\end{align*}

If $|a(t) - b (t)| \leq r:$
\begin{align*}
&a(t+1) = \frac{\tau_o - 1}{\tau_o}a(t) + \frac{\lambda(1-\omega)}{\tau_o}(pa(t)+(1-p)b(t)) + \frac{\lambda\omega}{\tau_o}a(0) + \frac{1-\lambda}{\tau_o}h(e(t)) \\
&b(t+1) = \frac{\tau_o - 1}{\tau_o}b(t) + \frac{\lambda(1-\omega)}{\tau_o}(pa(t)+(1-p)b(t)) + \frac{\lambda\omega}{\tau_o}b(0) + \frac{1-\lambda}{\tau_o}h(e(t))\\
&\implies a(t+1) - b(t+1) = \frac{\tau_o - 1}{\tau_o}(a(t) - b(t)) + \frac{\lambda\omega}{\tau_o}(a(0) - b(0)) \geq 0 \\
&\implies a(t+1) \geq b(t+1)
\end{align*}

\subsection{Proof of Proposition 2}
\label{appA2}
Let $\frac{\lambda\omega P(0)}{1 - \lambda(1-\omega)} < r$. First we want to show that $P(t)$ eventually is lower than r and that once this occurs it is lower than r for all future t. From our assumption we can state that $\exists  \ \epsilon > 0$ such that $\frac{\lambda\omega P(0)}{1 - \lambda(1-\omega)} + \epsilon < r$. If $\exists$ no $t$ such that $P(t) \leq r$ then $\forall \ t$ we can write:
\begin{align*}
P(t) &= (\frac{\tau_o - 1 + \lambda(1-\omega)}{\tau_o})^tP(0) + \frac{\lambda\omega P(0)}{\tau_o}\sum_{n=0}^{t-1}(\frac{\tau_o - 1 + \lambda(1-\omega)}{\tau_o})^n \\
&= (\frac{\tau_o - 1 + \lambda(1-\omega)}{\tau_o})^t(P(0) - \frac{\lambda\omega P(0)}{1 - \lambda(1-\omega)}) + \frac{\lambda\omega P(0)}{1 - \lambda(1-\omega)}
\end{align*}

Since $|\frac{\tau_o - 1 + \lambda(1-\omega)}{\tau_o}| < 1$ therefore $\lim_{t \xrightarrow{} \infty}P(t) = \frac{\lambda\omega P(0)}{1 - \lambda(1-\omega)}$. Therefore for $\epsilon$ defined above $\exists \ T$ such that $P(T) - \frac{\lambda\omega P(0)}{1 - \lambda(1-\omega)} < \epsilon$ which would imply $P(T) < \frac{\lambda\omega P(0)}{1 - \lambda(1-\omega)} + \epsilon < r$ which is a contradiction. Therefore $\exists \ T$ such that $P(T) \leq r$. Therefore $\forall \ t$:
\begin{align*}
p(t+T) &= (1 - \frac{1}{\tau_o})^tp(T) + \frac{\lambda\omega P(0)}{\tau_o}\sum_{n=0}^{t-1}(1-\frac{1}{\tau_o})^n \\
&=(1 - \frac{1}{\tau_o})^tp(T) + \lambda\omega P(0)(1 - (1-\frac{1}{\tau_o})^t) \\
&<(1 - \frac{1}{\tau_o})^tr + r(1 - (1-\frac{1}{\tau_o})^t) = r
\end{align*}

Therefore $\lim_{t\xrightarrow{} \infty} P(t) = \lim_{t\xrightarrow{} \infty}(1 - \frac{1}{\tau_o})^tP(T) + \lambda\omega P(0)(1 - (1-\frac{1}{\tau_o})^t) = \lambda\omega P(0)$. Let $\frac{\lambda\omega P(0)}{1 - \lambda(1-\omega)} > r$. We want to show that P(t) is always above r. For contradiction assume $\exists \ T$ such that $P(T) \leq r$. Let $T$ be the smallest such $T$. This implies that:
\begin{align*}
P(T) &= (\frac{\tau_o - 1 + \lambda(1-\omega)}{\tau_o})^t(P(0) - \frac{\lambda\omega P(0)}{1 - \lambda(1-\omega)}) + \frac{\lambda\omega P(0)}{1 - \lambda(1-\omega)} \\
&> (\frac{\tau_o - 1 + \lambda(1-\omega)}{\tau_o})^t(P(0) - \frac{\lambda\omega P(0)}{1 - \lambda(1-\omega)}) + r > r
\end{align*}

Which is a contradiction. Therefore $P(t) > r  \ \forall  \ t$, which implies that \\ $\lim_{t\xrightarrow{}\infty}P(t) = \lim_{t\xrightarrow{}\infty}(\frac{\tau_o - 1 + \lambda(1-\omega)}{\tau_o})^t(P(0) - \frac{\lambda\omega P(0)}{1 - \lambda(1-\omega)}) + \frac{\lambda\omega P(0)}{1 - \lambda(1-\omega)} = \frac{\lambda\omega P(0)}{1 - \lambda(1-\omega)}$.

\subsection{Convergence to consensus in non-prejudiced populations}
\label{appA3}
In non-prejudiced populations, if $P(0) > r$ then $P(t+1) = a(t+1) - b(t+1) = (1 - \frac{1}{\tau_o} + \frac{\lambda}{\tau_o})(a(t) - b(t)) = \frac{\tau_o - 1 + \lambda}{\tau_0}P(t)$. This implies that $P(t) = (\frac{\tau_o - 1 + \lambda}{\tau_0})^tP(0)$. Note that since $\tau_o > 1$ and $\lambda < 1$, $|\frac{\tau_o - 1 + \lambda}{\tau_o}| < 1$. Therefore $P(t) \leq r \iff (\frac{\tau_o - 1 + \lambda}{\tau_o})^t P(0) \leq r \iff t \geq \frac{ln(r/p(0))}{ln(\frac{\tau_o - 1 + \lambda}{\tau_o})}$. This means that no matter the initial polarization, the two polarized groups are within the confidence bound of each other at $T = \lfloor\frac{ln(r/p(0)}{ln(\frac{\tau_o - 1 + \lambda}{\tau_o})} + 1\rfloor$.

\section{}
\label{appB}

\subsection{Proof of Proposition 3}
\label{appB1}
If $c(t) = c(t+1)$ then $c(t) = \frac{\tau_o - 1 + \lambda}{\tau_o}c(t) + \frac{1 - \lambda}{\tau_o}sign(-\beta + e(t)) \implies c(t) = sign(-\beta + e(t))$. If $e(t) = e(t+1)$ then $e(t) = (1 - \gamma)e(t) + u(c(t)) \implies e(t) = \frac{1}{\gamma}u(c(t))$. Therefore, $c(t) = sign(-\beta + \frac{\alpha}{2\gamma}(1 - c(t))$. Since the sign function can only take on values of -1, 0, or 1 there are only 3 possible fixed points. 

If $c(t) = 1$ then $1 = sign(-\beta)$ which implies that $\beta < 0$. When c(t) = 1, then $e(t) = \frac{\alpha}{2\gamma}(1 - 1) = 0$. The dynamics when $\beta < 0$ reduce to $c(t+1) = \frac{\tau_o - 1 + \lambda}{\tau_o}c(t) + \frac{1 - \lambda}{\tau_o}$, $e(t+1) = (1-\gamma)e(t) + \frac{1}{2}\alpha(1 - c(t))$ whose Jacobian at $(c,e) = (1, 0)$ is: A = $\begin{pmatrix} \frac{\tau_o - 1 + \lambda}{\tau_o} & 0 \\ -\frac{1}{2}\alpha & 1 - \gamma \end{pmatrix}$. The eigenvalues of A are $\frac{\tau_o - 1 + \lambda}{\tau_o}$ and $1 - \gamma$ which are both within the unit circle, implying that $(c, e) = (1, 0)$ is a stable fixed point when $\beta < 0$.

If $c(t) = -1$ then $-1 = sign(-\beta + \frac{\alpha}{\gamma}) \implies \beta > \frac{\alpha}{\gamma}$. When $c(t) = -1$, then $e(t) = \frac{\alpha}{2\gamma}(1-(-1)) = \frac{\alpha}{\gamma}$. The Jacobian at $(c, e) = (-1, \frac{\alpha}{\gamma})$ is the same as the case above, A = $\begin{pmatrix} \frac{\tau_o - 1 + \lambda}{\tau_o} & 0 \\ -\frac{1}{2}\alpha & 1 - \gamma \end{pmatrix}$. Therefore, $(c, e) = (-1, \frac{\alpha}{\gamma})$ is a stable fixed point when $\beta>\frac{\alpha}{\gamma}$.

\subsection{Fixed points in prejudiced populations}
\label{appB2}
When $\frac{\lambda\omega P(0)}{1 - \lambda(1-\omega)} > r$, the steady-state conditions $a(t+1)=a(t), b(t+1)=b(t)$ can be solved since the polarized groups are independent at equilibrium. When $\frac{\lambda\omega P(0)}{1 - \lambda(1-\omega)} \leq r$, we are able to separate the dynamics of the two groups by assuming that the groups are separated by $\lambda\omega P(0)$, based on the results of Proposition 2. The expressions were simplified by setting $h(e(t)) = 1$ in the mitigative region and $h(e(t)) = -1$ in the non-mitigative regions. Similarly, the expressions for the environment at equilibrium were evaluated by solving the steady-state condition $e(t+1) = e(t)$ and substituting in the fixed point opinion profile. The oscillation centers were estimated by substituting $e(t) = \beta$ into the expression for the environment and then solving for $a(t)$ and $b(t)$.

When $\frac{\lambda\omega P(0)}{1 - \lambda(1-\omega)} > r$, the long-term behavior in the mitigative and non-mitigative region is:
\begin{align*}
a_m(t) &= qa(0) + (1-q) \ &&a_n(t) = qa(0) - (1-q) \\
b_m(t) &= qb(0) + (1-q) \ &&b_n(t) = qb(0) - (1-q) \\
e_m(t) &= \frac{\alpha}{2\gamma}q(1-pa(0) - (1-p)b(0)) \\ e_n(t) &= \frac{\alpha}{2\gamma}(2-q(1 + pa(0) + (1-p)b(0)))
\end{align*}
The environment and opinions in the cyclic region will oscillate around:
\begin{align*}
a_c(t) &\approx 1 - \beta\frac{2\gamma}{\alpha} + (1-p)qP(0) \\
b_c(t) &\approx 1 - \beta\frac{2\gamma}{\alpha} - pqP(0) \\
e_c(t) &\approx \beta
\end{align*}
When $\frac{\lambda\omega P(0)}{1 - \lambda(1-\omega)} \leq r$, the long-term behavior in the mitigative and non-mitigative region is:
\begin{align*}
a_m(t) &= qa(0) + (1-q) - \lambda(1-\omega)(1-p)qP(0) \\
b_m(t) &= qb(0) + (1-q) + \lambda(1-\omega)pqP(0) \\
e_m(t) &= \frac{\alpha}{2\gamma}q(1-pa(0) - (1-p)b(0)) \\
a_n(t) &= qa(0) - (1-q) -\lambda(1-\omega)(1-p)qP(0) \\
b_n(t) &= qb(0) - (1-q) + \lambda(1-\omega)pqP(0) \\
e_n(t) &= \frac{\alpha}{2\gamma}(2-q(1 + pa(0) + (1-p)b(0)))
\end{align*}
The environment and opinions in the cyclic region will oscillate around:
\begin{align*}
a_c(t) &\approx 1 - \beta\frac{2\gamma}{\alpha} + (1-p)\lambda\omega P(0) \\
b_c(t) &\approx 1 - \beta\frac{2\gamma}{\alpha} - p\lambda\omega P(0) \\
e_c(t) &\approx \beta
\end{align*}

\section{}
\label{appC}

\subsection{Distance from the objective best response}
\label{appC1}
To analyze how far the population is from the 'truth' we can define the distance from the truth $T$ as $d_T = \frac{1}{N}\sum_{i=1}^N|o_i(t) - T|$. In our polarized model this function reduces to $d_T = p|a(t) - T| + (1-p)|b(t) - T|.$ We can only investigate the truth when in the non-mitigative and mitigative regions as the truth is well-defined in those regions. In the cyclic region, the truth is constantly oscillating between mitigation and non-mitigation since the environmental cost continually oscillates above and below the cost of mitigating. We must also separate our analysis dependent on whether $\frac{\lambda\omega P(0)}{1 - \lambda(1-\omega)} > r$.

In the mitigative region $T=1$. When $\frac{\lambda\omega P(0)}{1 - \lambda(1-\omega)} > r$, we have that $a_m(t) = qa(0) + (1-q)$ and $b_m(t) = qb(0) + (1-q)$. Substituting these into $d_T$ we find that: $d_T = q(p|a(0) - 1| + (1 - p)|b(0) - 1|)$. In the non-mitigative region $T=-1$, $a_n(t) = qa(0) - (1-q)$ and $b_n(t) = qb(0) - (1-q)$, which implies that $d_T = q(p|a(0) + 1| + (1 - p)|b(0) + 1|)$. In both cases, the distance from the truth increases linearly with the proportion of prejudice, no matter what those prejudice's are. Additionally, a polarized population converges onto the truth if and only if there is no prejudice.

In the case that $\frac{\lambda\omega P(0)}{1 - \lambda(1-\omega)} \leq r$, the expressions for the long-term behavior of $a(t)$ and $b(t)$ change. In the mitigative region $T=1$, $a_m(t) = qa(0) + (1-q) - \lambda(1-\omega)(1-p)qP(0)$, $b_m(t) = qb(0) + (1-q) + \lambda(1-\omega)pqP(0)$. This implies $d_T = q(p|a(0)-\lambda(1-\omega)(1-p)P(0) - 1| + (1-p)|b(0)+\lambda(1-\omega)pP(0) -1|)$. In the non-mitigative region $d_T = q(p|a(0)-\lambda(1-\omega)(1-p)P(0) + 1| + (1-p)|b(0)+\lambda(1-\omega)pP(0) +1|)$. In both cases, if the amount of social susceptibility is kept as a constant, the distance to the truth increases linearly with the proportion of prejudice, no matter what those prejudice's are. As in the previous case, a polarized population converges onto the truth if and only if there is no prejudice.

\section{}
\label{appD}
\setcounter{figure}{0}

\begin{figure}[H]
    \includegraphics[width=1\linewidth]{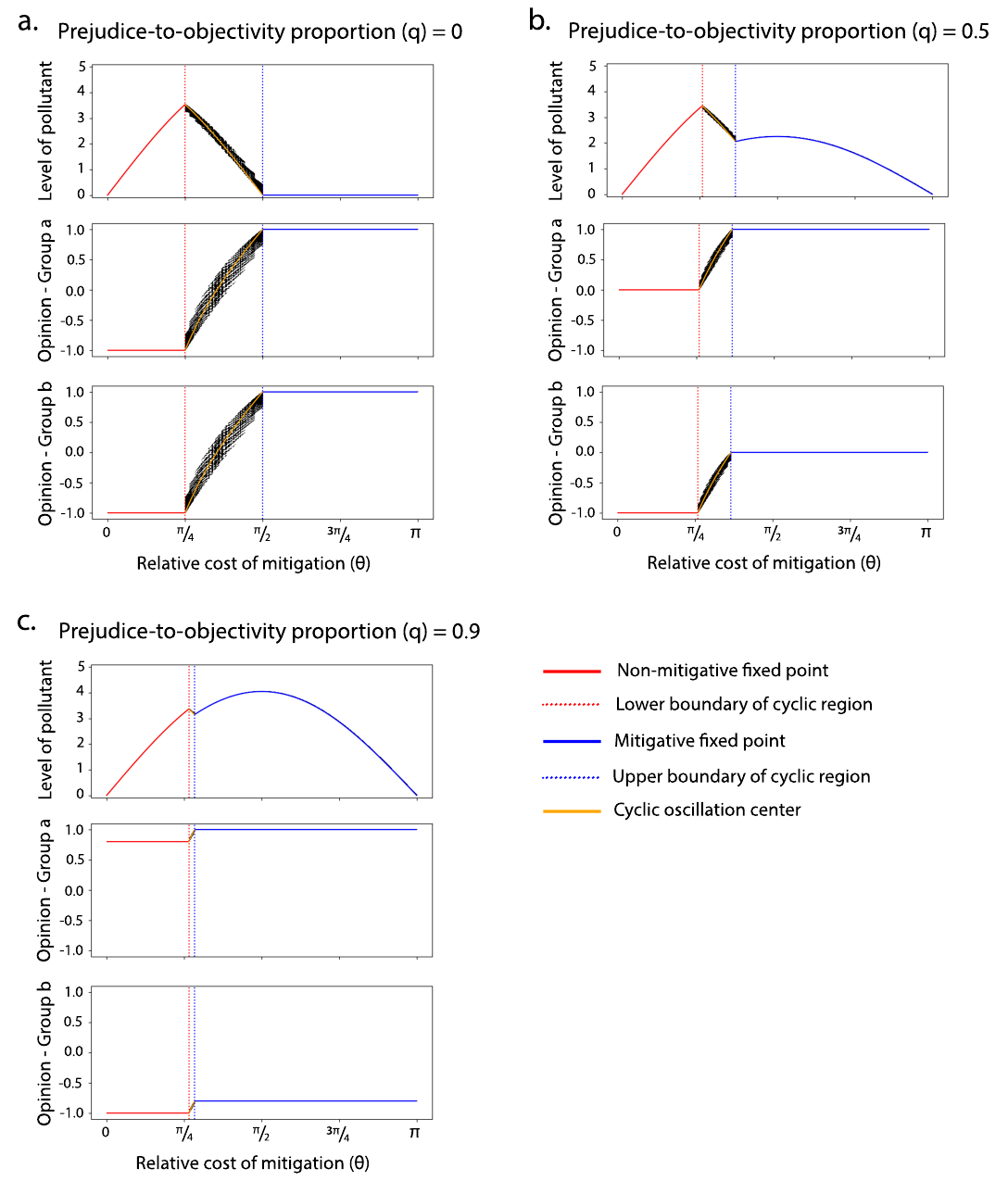}
    \caption{\textbf{Validation of fixed point expressions.} Overlay of the derived expressions for fixed points and long-term behavior with numerical bifurcation diagrams over the relative cost of mitigation ($\theta$). Expressions show alignment in populations with (a) no prejudice ($q = 0$), (b) equal prejudice and objectivity ($q = 0.5$), and (c) high levels of prejudice ($q=0.9$).}
    \label{FigureD.1}
\end{figure}

\begin{figure}[H]
    \centering
    \includegraphics[width=0.75\linewidth]{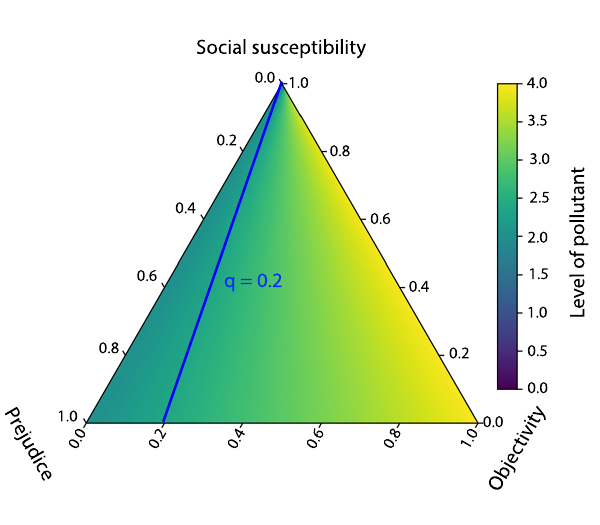}
    \caption{\textbf{Effect of influence coefficients on the level of pollutant.} Heatmap of the long-term level of pollutant over varying strengths of social susceptibility, prejudice, and objectivity of the population. Simulations were performed at $\alpha = 0.8, \beta = 5, \gamma = 0.2$.}   
    \label{FigureD.2}
\end{figure}

\end{document}